\documentclass[aps,twocolumn,nofootinbib,superscriptaddress]{revtex4}

\newcommand{\pt}{p_{\rm T}}
\newcommand{\sq}{\sqrt{s_{\rm NN}}}
\newcommand{\vp}{v^\prime}
\newcommand{\vv}{v_{\rm 1}}
\newcommand{\vvp}{v_{\rm 1}^\prime}
\newcommand{\vvpa}{v_{\rm 1,1-3}^\prime}
\newcommand{\vvpb}{v_{\rm 1,4}^\prime}
\newcommand{\vvpc}{v_{\rm 1,5}^\prime}

\usepackage{float}
\usepackage{amsmath}
\usepackage{graphicx}
\usepackage[colorlinks,linkcolor=blue,urlcolor=blue,citecolor=blue]{hyperref}

\begin{document}
\title{Coalescence sum rule and the electric charge- and strangeness-dependences of directed flow in heavy ion collisions} 

\author{Kishora Nayak}
\affiliation{Department of Physics, Panchayat College, Sambalpur University, Bargarh 768028, Odisha, India}
\affiliation{Key Laboratory of Quark \& Lepton Physics (MOE) and
  Institute of Particle Physics, Central China Normal University,
  Wuhan, 430079, China} 
\author{Shusu Shi}
\affiliation{Key Laboratory of Quark \& Lepton Physics (MOE) and 
  Institute of Particle Physics, Central China Normal University, 
  Wuhan, 430079, China} 
\author{Zi-Wei Lin}
\email{linz@ecu.edu}
\affiliation{Department of Physics, East Carolina University,
  Greenville, NC, 27858, USA}

\begin{abstract}
The rapidity-odd directed flows ($v_{\rm 1}$) of identified hadrons are
expected to follow the coalescence sum rule
when the created matter is initially in parton degrees of freedom  
and then hadronizes through quark coalescence. 
A recent study has considered the $v_{\rm 1}$ of produced hadrons that
do not contain $u$ or $d$ constituent quarks. 
It has constructed multiple hadron sets with a small 
mass difference but given difference in electric charge $\Delta q$ and  
strangeness $\Delta S$ between the two sides, where a nonzero and
increasing $\Delta v_{\rm 1}$ with $\Delta q$ has been proposed to be
a consequence of electromagnetic fields. In this study, we
examine the consequence of coalescence sum rule on the $\Delta v_{\rm
  1}$ of the hadron sets in the absence of electromagnetic
fields. We find that in general $\Delta v_{\rm 1} \neq 0$ for a hadron
set with nonzero $\Delta q$ and/or $\Delta S$ due to potential
$v_{\rm 1}$ differences between $\bar u$ and $\bar d$ and between $s$
and $\bar s$ quarks.  We further propose methods to extract the 
coefficients for the $\Delta q$- and $\Delta S$-dependences of the
direct flow difference, where a nonzero constant term
would indicate the breaking of the coalescence sum rule. 
The extraction methods are then demonstrated with transport model
results.
\end{abstract}


\maketitle

\section{Introduction}

The properties of the quark-gluon plasma produced in relativistic
heavy ion collisions can be studied with the directed flow
($\vv$)~\cite{Li:1998ze,Stoecker:2004qu,Nara:2016phs,Luo:2020pef}. For
example, $\vv$ is found to be a sensitive probe of the equation of
state of the produced  matter~\cite{nuclPassage1,nuclPassage2}, and
$\vv$ of heavy flavors~\cite{Das:2016cwd} is expected to be sensitive
to the strong electromagnetic field in the early stage of noncentral
heavy ion collisions.

The coalescence sum rule is often found
to describe well the relations of anisotropic flows of different hadron 
species in heavy ion collisions at high
energies~\cite{Molnar:2003ff,ncqSTAR,ncqSTAR2,ncqPHENIX1,ncqPHENIX2,ncqALICE}. 
For collisions where the dynamics of
anisotropic flows is dominated by parton interactions,  
quark coalescence relates the hadron flow directly to the flows of the
hadron's constituent quarks~\cite{Lin:2002rw,Molnar:2003ff,Lin:2003jy}. 
When the constituent quarks in a hadron are comoving with
each other and the quark coalescence probability is small, 
the hadron elliptic flow $v_2$ follows the coalescence sum rule at
leading order~\cite{Molnar:2003ff,Lin:2003jy,Lin:2009tk}. The same
formulation can  be extended to the directed flow. When we
neglect the mass difference of the constituent
quarks~\cite{Lin:2003jy}, the coalescence sum rule is simply given
by~\cite{Molnar:2003ff} 
\begin{equation}
v_n^H(\pt^H) =  \sum_j  v_{n,j} (\pt), {\rm~with~}\pt^H=N_{cq}\,\pt.
\label{csr}
\end{equation}
In the above, $n=1$ for $\vv$ and $n=2$ for $v_2$, 
$v_{n,j} (\pt)$ represents the flow $v_n$ of constituent quark $j$
at the quark transverse momentum $\pt$, while $N_{cq}$ is the number
of constituent quarks (NCQ) of the hadron species $H$. Furthermore, if
the quark $v_n(\pt)$ is the same for each constituent quark 
of hadron species $H$, Eq.\eqref{csr} reduces to the most used form of
the NCQ scaling: $v_n^H(N_{cq} \,\pt)=  N_{cq} \, v_n(\pt)$.

It has been proposed~\cite{starPRL} that the direct flows of hadrons whose
constituent quarks are all produced quarks can be properly combined to
better test the coalescence sum rule. In contrast to produced quarks, hadrons
containing $u$ and/or $d$ quarks get contributions from slowed-down
(or transported) $u$ and $d$ quarks in the incoming
nuclei~\cite{transport1,myPRC_AMPT}, which complicate the flow
analysis. Our study here has been motivated by a recent
study~\cite{referAshik}, which further considered the $\vv$ difference
of various combinations of hadron sets consisting of seven 
produced hadron species: $K^-,\phi,\bar p, \bar \Lambda, {\bar \Xi}^+,
\Omega^-$, and ${\bar \Omega}^+$. For example, one of the combinations 
is $\vv [{\bar \Lambda}]-(\vv[\phi]/2+2\vv[\bar p]/3)$. That study
focused on the dependence of the $\vv$ difference on the electric charge
difference $\Delta q$ and the strangeness difference $\Delta S$ of the
hadron set combinations.  
A nonzero $\vv$ difference at nonzero $\Delta q$
was considered as the breaking of the coalescence sum rule 
and proposed to be a consequence of the electromagnetic
fields~\cite{referAshik,STAR:2023wjl}, especially if the $\vv$ difference
increases with $\Delta q$. The study also recognized the need for
further investigation if a systematic dependence of the $\vv$
difference on $\Delta S$ is observed~\cite{referAshik}.

In this study, we examine in detail the $\vv$
difference of various combinations of these seven hadron species. 
Note that the $v_1$ throughout this study refers to the rapidity-odd 
directed flow, although $\vv$ contains both rapidity-odd
and rapidity-even components where the rapidity-even directed flow
originates from event-by-event fluctuations. 
In addition, since we only consider light quarks, which constituent
masses are not too different, we neglect the effect of different 
quark masses on the coalescence sum rule~\cite{Lin:2003jy} and thus
start the analysis from Eq.\eqref{csr}.  The paper is organized as follows. 
In Sec.~\ref{csrv1}, we derive the coalescence sum rule relationships
between the $\vv$ difference of each hadron set and the quark
$\vv$. In Sec.~\ref{coeff}, we present two methods to 
extract the dependences of the $\vv$ difference on the electric charge
difference $\Delta q$ and the strangeness difference $\Delta S$, and
in Sec.~\ref{ampt} we demonstrate the extraction methods with the 
numerical $\vv$ results from a multi-phase transport (AMPT) model.
Finally, we summarize in Sec.~\ref{conclusions}.

\section{Coalescence sum rule relations for the $\vv$ difference of a hadron set}
\label{csrv1}

\setlength{\tabcolsep}{6pt} % Default value: 6pt
\renewcommand{\arraystretch}{1.5} % Default value: 1
\begin{table*}[ht]
\begin{tabular}{cccccc}
%\toprule
\cline{1-6}
Set \# &  $\Delta q_{ud}$ & $\Delta S$ & $\Delta q$  & L (left side) & R (right side)\\
\cline{1-6}%\midrule 
   1    &       0       &   0  &  0  &
$\vv[K^-(\bar{u}s)]+\vv [\overline{\Lambda}(\bar{u}\bar{d}\bar{s})]$  
& $\vv[\bar{p}(\bar{u}\bar{u}\bar{d})]+\vv[\phi(s\bar{s})]$ \\ 
   2    &      0      &  0  &  0   &   
$\vv\big[\overline{\Lambda}(\bar{u}\bar{d}\bar{s})\big]$
                                             & $\frac{1}{2}\vv[\bar{p}(\bar{u}\bar{u}\bar{d})] + \frac{1}{2} \vv [\overline{\Xi}^{+}(\overline{d}\bar{s}\bar{s})]$  \\ 
   3    &     0      &  0  &  0  &    $\frac{1}{3}\vv[\Omega^{-}(sss)]+\frac{1}{3}\vv[\overline{\Omega}^{+}(\bar{s}\bar{s}\bar{s})]$    & $\vv[\phi(s\bar{s}) ]$   \\
\cline{1-6} 
   4  &     0       &  1 &    1/3  &  
$\frac{1}{2} \vv[\phi(s\bar{s})]$ & $\frac{1}{3}\vv[\Omega^{-}(sss)]$ \\  
\cline{1-6}
   5A    &      1/3      &  1 &  2/3  &   $\frac{1}{2}\vv[\phi(s\bar{s})]+\frac{1}{3}\vv[\bar{p}(\bar{u}\bar{u}\bar{d})]$  & $\vv\big[K^{-}(\bar{u}s)\big] $ \\         
   5B    &      1/3      &  1 &  2/3  &    
$\vv \big[\overline{\Lambda}(\bar{u}\bar{d}\bar{s})\big]$
              & $\frac{1}{2}\vv[\phi(s\bar{s})]+\frac{2}{3}\vv[\bar{p}(\bar{u}\bar{u}\bar{d})]$
  \\    
\cline{1-6} %\bottomrule 
\end{tabular}%} 
\caption{List of several hadron sets, where the left side and right side have 
the same total number of $\bar u$ and $\bar d$ quarks and the same
total number of $s$ and $\bar s$ quarks (after including the
weights). $\Delta q$, $\Delta q_{ud}$ and $\Delta S$ represent  
the difference in the electric charge, electric charge from $\bar u$
and $\bar d$ quarks, and strangeness number, respectively, between
the two sides. Note that sets 1 to 4 and set 5A are independent of
each other, while set 5B is not independent of them.}  
\label{table1}
\end{table*}

In this study, we only consider produced hadrons whose constituent
quarks consist of  $\bar{u}$, $\bar{d}$, $s$ and $\bar{s}$ quarks. 
Table~\ref{table1} lists several such hadron sets, 
where for each combination the left side and the right side have 
the same total number of $\bar u$ and $\bar d$ quarks and the same
total number of $s$ and $\bar s$ quarks (after including the
weighting factors). 
For a given hadron set, let $N_i^L$ and $N_i^R$ be the total number of
constituent quarks of flavor $i$ in each hadron multiplied by the
weighting factor of the hadron on the left side and right side,
respectively.  We then write 
\begin{equation}
\Delta N_i \equiv N_{i}^{L}-N_{i}^{R}
\label{ni}
\end{equation}
as the difference of $N_i$ between the two sides. 
Then each hadron set in Table~\ref{table1} satisfies the following relations: 
\begin{equation}
\Delta N_{\bar u} + \Delta N_{\bar d}= 0, ~\Delta N_{s} + \Delta N_{\bar s} = 0.
\label{dn}
\end{equation}
For example, set 5A has $N_{\bar u}^L=2/3$, $N_{\bar d}^L=1/3$,
$N_s^L=N_{\bar s}^L=1/2$, $N_{\bar u}^R=1$, and $N_s^R=1$.
Similar to Eq.\eqref{ni}, we can define 
the differences of the total electric charge
in $\bar u$ and $\bar d$ quarks ($q_{ud}$), the total strangeness $S$, 
and the total electric charge $q$, between the two sides as 
\begin{eqnarray}
\Delta q_{ud}& \equiv &q_{ud}^L-q_{ud}^R=\Delta N_{\bar  d}, \nonumber \\
\Delta S&\equiv &S^L-S^R=2\Delta N_{\bar s}, \nonumber \\
\Delta q&\equiv &q^L-q^R =\Delta q_{ud}+\frac{1}{3} \Delta S,
\label{dnqs}
\end{eqnarray}
respectively. 
The values of $\Delta q_{ud}$, $\Delta S$, and $\Delta q$ for each
hadron set are given in Table~\ref{table1}, where the left side and
right side are shown with the constituent quark content and the
weighting factor of each hadron. Because of Eq.\eqref{dn}, the mass
difference (after including the weighting factors) between the two
sides is small for most of these hadron sets. Note that sets 1, 2,
and 3 each have identical constituent quark content on the 
left and right sides and thus satisfy $\Delta q_{ud}=\Delta
q=\Delta S=0$. On the other hand, sets 4, 5A and 5B each have a
nonzero charge difference and/or a nonzero strangeness difference
between the two sides. One can show that the conditions
of Eq.\eqref{dn} lead to the following general hadron set:
\begin{eqnarray}
&&a_1\, K^-+a_2\,\phi+a_3\,\bar p+a_4 \,\bar \Lambda
-(a_1+3a_3+2a_4) \,{\bar \Xi}^+ \nonumber \\
&&+a_5 \,\Omega^-+\left ( \frac{a_1}{3}-\frac{2a_2}{3} +2a_3+a_4 -a_5\right )
   {\bar \Omega}^+=0, 
\end{eqnarray}
where $a_i$ are arbitrary constants; as a result, there are only five
sets of  independent hadron sets \footnote{We realized that there are
  only five independent hadron sets under the constraint of
  Eq.\eqref{dn} in August 2021.}\cite{referAshik}. 
Sets 1 to 4 and 5A in Table~\ref{table1} give one example of the five
independent sets; so do sets 1 to 4 and 5B. However, sets 1, 5A, and 5B
are not independent of each other, since the $\vv$ difference
between the two sides of set 5B can be written as  that of set 5A plus
that of set 1. With sets 1 to 4 and 5A (or 5B) 
in Table~\ref{table1}, one can construst all the hadron sets of 
earlier studies~\cite{referAshik,STAR:2023wjl}.

We now apply the coalescence sum rule in Eq.\eqref{csr} to 
evaluate the difference between the $\vv$ from two sides of
a given hadron set. 
Since we neglect the mass difference of $u/d/s$ constituent quarks, 
the quarks coalescing to form a hadron have the same $\pt$. 
If we only consider quarks at a given $\pt$, then they will form
mesons at $\pt^M=2 \pt$ and (anti)baryons at $\pt^B=3  \pt$;  
this is why we have chosen the $\pt$ range as $[0,2]$ GeV$/c$ for mesons
and $[0,3]$ GeV$/c$ for (anti)baryons for the analysis of the model
calculations in Sec.~\ref{ampt}. The difference between the $\vv$
from two sides of a given hadron set is then given by
\begin{equation}
\Delta \vv \equiv \vv^L- \vv^R= \sum_{i} \Delta N_i \; v_{\rm 1,i}\; , 
\label{dv1}
\end{equation}
where $v_{\rm 1,i}$ represents the $v_1$ of quark flavor $i$ 
with $i \in \{ \bar u, \bar d, s, \bar s \}$ and we have skipped the $\pt$
argument in the $v_1(\pt)$ notations for brevity. 
Note that although the above relation is written for a given quark
$\pt$, it still applies when quarks are selected within a given $\pt$
range, in which case $v_{\rm 1,i}$ just represents the average $v_1$ of
quark flavor $i$ within that $\pt$ range. 
With Eqs.\eqref{dn}-\eqref{dnqs}, we further obtain
\begin{eqnarray}
&&\Delta \vv=( v_{1,\bar d} - v_{1,\bar u} )  \Delta q_{ud} 
+ \left ( \frac{v_{1,\bar s} - v_{1,s}}{2}  \right ) \Delta S \nonumber \\
&&= ( v_{1,\bar d} \!-\! v_{1,\bar u}) \Delta q
    \!+\! \left ( \frac{v_{1,\bar s} \!-\! v_{1,s}}{2}
    \!-\! \frac{v_{1,\bar d} \!-\! v_{1,\bar u}}{3} \right ) \Delta S.
\label{v1q}
\end{eqnarray}

$\vv$ observables such as those appearing in
Eqs.\eqref{dv1}-\eqref{v1q} are functions of the hadron rapidity
$y$. The rapidity-odd $\vv$ around mid-rapidity is often fit with a
linear function in rapidity, with the only parameter being the slope
($\vvp$). If we assume that the rapidity of a hadron formed 
by quark coalescence is the same as that of the coalescing quarks
(which have the same rapidity due to the comoving requirement),  
we can then take the derivative with respect to $y$ and obtain
\begin{equation}
\Delta \vvp = \sum_{i} \Delta N_i \; \vp_{\rm 1,i}\; , {\rm~with~}
\vvp \equiv \frac{d\vv}{dy} {\Big |}_{y=0}.
\label{dv1p}
\end{equation}
The above just relates the difference of the $\vv$ slope parameters
from two sides of a hadron set to the quark $\vv$ slope
parameters. We also have
\begin{eqnarray}
&&\Delta \vvp=( \vp_{1,\bar d} - \vp_{1,\bar u} )  \Delta q_{ud} 
+ \left ( \frac{\vp_{1,\bar s} - \vp_{1,s}}{2}  \right ) \Delta S \nonumber \\
&&= ( \vp_{1,\bar d} \!-\! \vp_{1,\bar u}) \Delta q
    \!+\! \left ( \frac{\vp_{1,\bar s} \!-\! \vp_{1,s}}{2}
    \!-\! \frac{\vp_{1,\bar d} \!-\! \vp_{1,\bar u}}{3} \right ) \Delta S.
\label{eqq}
\end{eqnarray}
Therefore, the difference of the $\vv$ slope parameters of a hadron
set depends linearly on both $\Delta q_{ud}$ and $\Delta S$, where the
corresponding coefficient is given by the difference of the
quark-level $\vv$ slope parameters. 
It is also clear that the interpretation of
the coefficients is simpler if we use $\Delta q_{ud}$ instead of
$\Delta q$ for the electric charge difference. 
When one assumes that $\bar u$ and $\bar d$ quarks have the same $\vv$ 
slope and that $s$ and $\bar s$ have the same $\vv$
slope~\cite{referAshik}, all the coefficients in Eq.\eqref{eqq} would
be zero. However, $\Delta \vvp \neq 0 $ in general according to the
coalescence sum rule when $\Delta q$ and/or $\Delta S$ is nonzero, 
which is the case for sets 4, 5A, and 5B in Table~\ref{table1}.

\section{Extracting coefficients for the $\Delta q$ and $\Delta S$ dependences} 
\label{coeff}

Since there are five independent sets, e.g., sets 1 to 4 and 5A, one 
will get five independent $\Delta \vvp$ data points from the
experimental measurement  (for a given event class of a given
collision system). One can then extract the $\Delta q$ and $\Delta S$
coefficients, which reflect the quark-level $\vv$ slope differences.
One way to extract the coefficients is to simply fit the 
five data points; this is the 5-set method. 
Alternatively, since sets 1 to 3 all have $\Delta q_{ud}=\Delta
q=\Delta S=0$,  we can combine these three data 
points into one and then fit three data points (the combined point
plus sets 4 and 5); this is the 3-set method.

For certain collision systems, the coalescence sum rule may not be
satisfied, e.g., if $\vv$ is not dominated by parton dynamics or the
flows are affected by other effects such as the electromagnetic field. 
Since Eq.\eqref{eqq} based on the coalescence sum rule gives
$\Delta \vvp=0$ for $\Delta q_{ud}=\Delta S=0$  
(and for $\Delta q=\Delta S=0$), we use the following modified
equations to fit the 5-set or 3-set $\Delta \vvp$ values:
\begin{eqnarray}
\Delta \vvp &=&c_0+c_q \Delta q_{ud}  + c_S \Delta S 
\label{fitqud} \\
&=& c_0^\star+c_q^\star\Delta q  + c_S^\star \Delta S.
\label{fitq}
\end{eqnarray}
This way, a nonzero value of the new intercept term $c_0$ or
$c_0^\star$ would mean the breaking of coalescence sum rule.
According to Eq.\eqref{eqq}, the coalescence sum rule predicts the 
following:
\begin{eqnarray}
&&c_0=c_0^\star=0, \nonumber \\
&&c_q=c_q^\star=\vp_{1,\bar d} - \vp_{1,\bar u}\; , \nonumber \\
&&c_S=\frac{\vp_{1,\bar s} - \vp_{1,s}}{2},  ~~c_S^\star=c_S-\frac{c_q}{3}.
\label{eqcoeff}
\end{eqnarray}

In the 3-set method, we combine the three $\Delta \vvp$ points (from
sets 1 to 3) into one point. Because these three data sets can have
very different statistical errors ($e_i$) or hadron counts, 
we average the central values of the three $\Delta \vvp$ data points
by using $1/e_i^2$ as the weight, and we calculate the 
statistical error of the combined data point as $1/\sqrt
{1/e_1^2+1/e_2^2+1/e_3^2}$. Let us denote the combined data point as
$\Delta \vvpa$; we also denote the data point from  
sets 4 and 5 (5A or 5B) as $\Delta\vvpb$ and $\Delta\vvpc$, respectively. 
Eq.\eqref{fitqud} then leads to $\Delta \vvpa =c_0, \Delta \vvpb =c_0 +
c_S, \Delta \vvpc =c_0+c_q/3 + c_S$. Therefore, the coefficients in
Eq.\eqref{fitqud} for the 3-set method are given by 
\begin{eqnarray}
c_0&=&\Delta \vvpa\; , ~~c_q=-3(\Delta \vvpb-\Delta \vvpc), \nonumber \\
c_S&=&-\Delta \vvpa+\Delta \vvpb.
\label{cqud3set}
\end{eqnarray}
Similarly, the coefficients in Eq.\eqref{fitq} for the 3-set method are given by
\begin{eqnarray}
c_0^\star&=&\Delta \vvpa\; ,~~c_q^\star=-3(\Delta \vvpb-\Delta \vvpc),
             \nonumber \\
c_S^\star &=&-\Delta \vvpa+2\Delta \vvpb-\Delta \vvpc.
\label{cq3set}
\end{eqnarray}

\section{Tests with a transport model}
\label{ampt}

We now use the AMPT model~\cite{ampt1} as an example to demonstrate
the $\vv$ analysis and extraction of the $\Delta q$ and $\Delta S$
coefficients.  We use the default version of the AMPT model to
simulate mid-central (10-50$\%$) Au+Au collisions at $\sq$ = 7.7,
14.5, 27, 54.4, and 200 GeV. The event centrality is determined from
the multiplicity of charged hadrons within the pseudorapidity range
$|\eta|<1/2$. For simplicity, we calculate $\vv$ with respect to the
reaction plane angle ($\Psi_{RP}$) as $\vv = \langle {\rm 
  cos}(\phi-\Psi_{RP})\rangle$, where $\phi$ is the azimuthal angle of
a hadron's momentum~\cite{flowDef1,flowDef3}. 

\begin{figure*}[htb]
\centering
\includegraphics[width=0.90\linewidth]{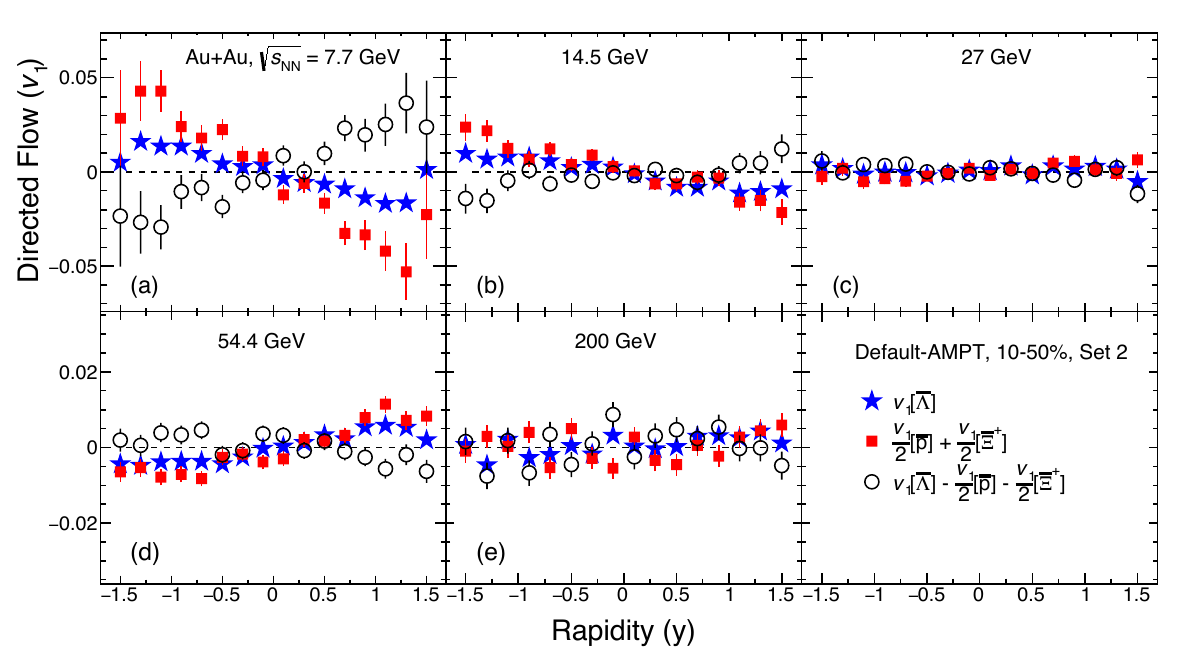}
\caption{Directed flows of hadron set 2 in Table~\ref{table1}:
  $\vv[\bar \Lambda]$, $\vv[{\bar p}]/2+\vv[{\bar \Xi}^{+}]/2$, and
  their difference as functions of rapidity from the default AMPT
  model for 10-50\% central Au+Au collisions at $\sq$ = 7.7,   14.5,
  27, 54.4 and 200 GeV.}
\label{fig1}
\end{figure*}

As an example, Fig.~\ref{fig1} shows the rapidity dependence of $\vv$
for hadron set 2, where $\vv^L=\vv[\bar \Lambda]$ and $\vv^R=\vv[{\bar
  p}]/2+\vv[{\bar \Xi}^{+}]/2$. 
We then fit their difference $\Delta \vv$ (circles) within $|y|<1.5$
at each energy with a rapidity-odd linear function of $y$ to obtain
the slope difference $\Delta \vvp$. Note that for hadron set 2 with 
$\Delta q_{ud}=\Delta q=\Delta S=0$, we expect $\Delta \vvp=0$ from
Eq.\eqref{eqq}. However, this is not the case for the default-AMPT
model results at low energies in Fig.~\ref{fig1}. 

\begin{figure*}[htb]
\centering
\includegraphics[width=0.90\linewidth]{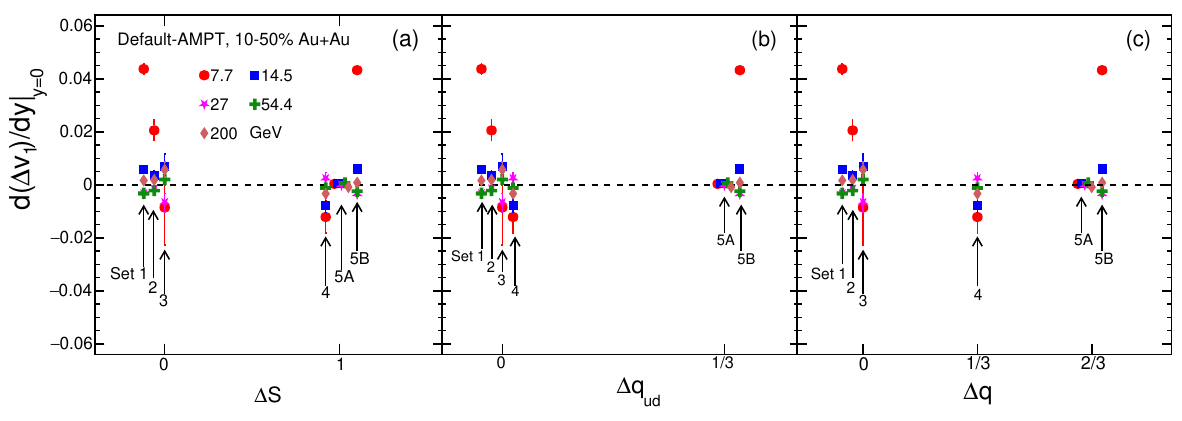}
\caption{The difference of the $\vv$ slopes at
  mid-rapidity for each hadron set in Table~\ref{table1} versus (a) $\Delta
  S$, (b) $\Delta q_{ud}$, and (c) $\Delta q$ from the default  AMPT
  model for 10-50$\%$ central Au+Au collisions at several energies.  
Data points at the same horizontal value are often slightly shifted
horizontally for better visibility.}
\label{fig2}
\end{figure*} 

\begin{figure*}[htb] 
\centering
\includegraphics[width=0.90\linewidth]{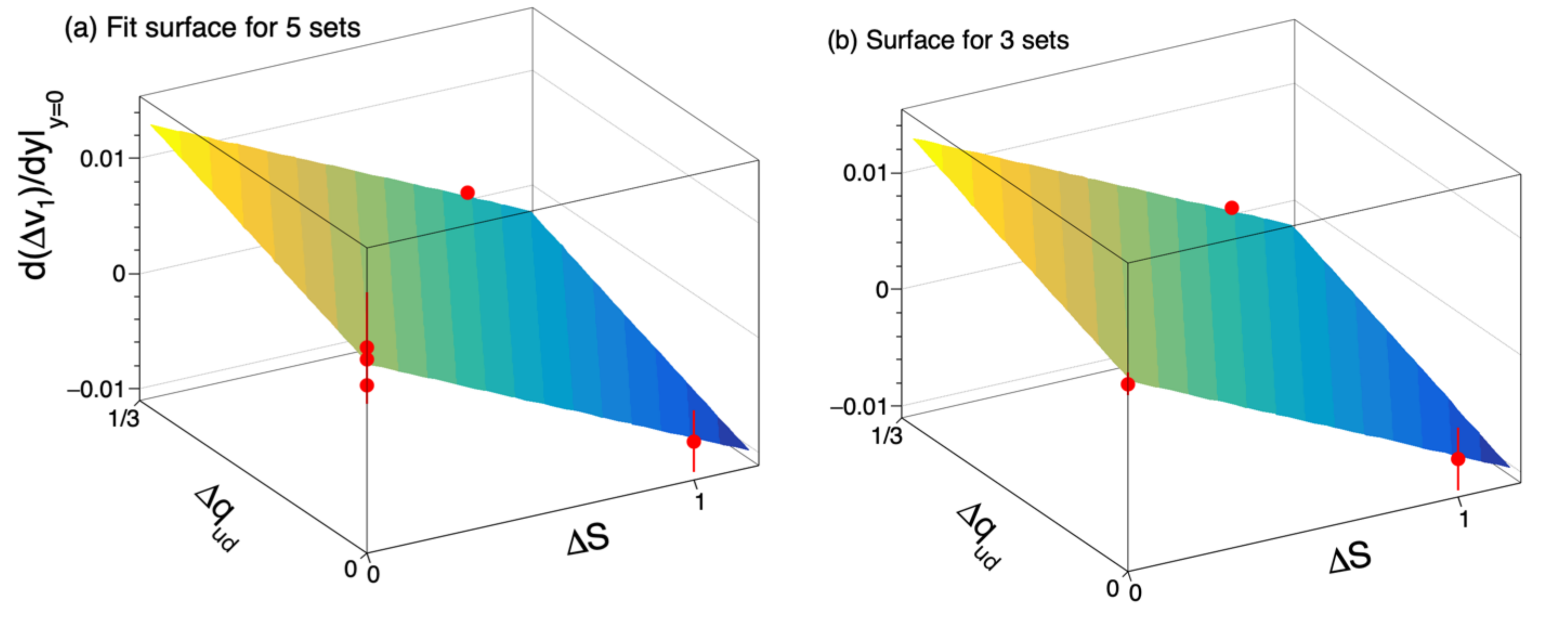}
\caption{The fit plane to extract the $\Delta \vvp$ dependences on
  $\Delta q_{ud}$ and $\Delta S$ as shown in Eq.\eqref{fitqud} 
using (a) the 5-set method and (b) the 3-set method.
The data points, corresponding to sets 1 to 4 and set 5A in
Table~\ref{table1}, come from the default AMPT model for mid-central
(10-50$\%$) Au+Au collisions at $\sq$ = 14.5 GeV.} 
\label{fig3}
\end{figure*}

Figure~\ref{fig2} shows the slope difference $\Delta \vvp$ of each set
at the five energies as functions of (a) $\Delta S$, 
(b) $\Delta q_{ud}$, and (c) $\Delta q$. 
Since $\Delta \vvp$ depends linearly on both $\Delta q$ and $\Delta
S$, one cannot determine the coefficient $c_q$ (or $c_S$) by simply 
performing a one-dimensional linear fit of the $\Delta q$ plot such as
Fig.~\ref{fig2}(b) (or the $\Delta S$ plot such as
Fig.~\ref{fig2}(a))~\cite{STAR:2023wjl}. 
Note that a one-dimensional linear fit as a function of $\Delta q$
performed at the same $\Delta S$ value~\cite{referAshik} would be
better. Here, we propose to extract the
$c_q$ and $c_S$ coefficients by describing the $\Delta \vvp$ data with
a two-dimensional plane (over the $\Delta q$-$\Delta S$ space). 
We can use the 5-set method by fitting five independent data
points with the relation of Eq.\eqref{fitqud}. 
As a demonstration, Fig.~\ref{fig3}(a) shows the fitting of five data
points (from sets 1 to 4 and set 5A) from the AMPT model at $\sq$
=14.5 GeV with the 5-set method.
Alternatively, we can use the 3-set method, where we fit the combined
data point for sets 1 to 3 and the data points from set 4 and set 5A
(or 5B). This is demonstrated in Fig.~\ref{fig3}(b), where the data
point at $\Delta q_{ud}=\Delta S=0$ represents an average of the
three corresponding data points shown in Fig.~\ref{fig3}(a) (from the
three hadrons sets with identical constituent quark content on the two
sides). The resultant coefficients obtained from the 5-set method and
the  3-set method are practically the same, as we can see from the
almost  identical planes in Figs.~\ref{fig3}(a) and (b). 
On the other hand, the 3-set method has an advantage in that the 
coefficients can be determined by Eq.\eqref{cqud3set} or
Eq.\eqref{cq3set} without the need to perform a fit.

\begin{figure*}[htb]
\centering
\includegraphics[width=0.90\linewidth]{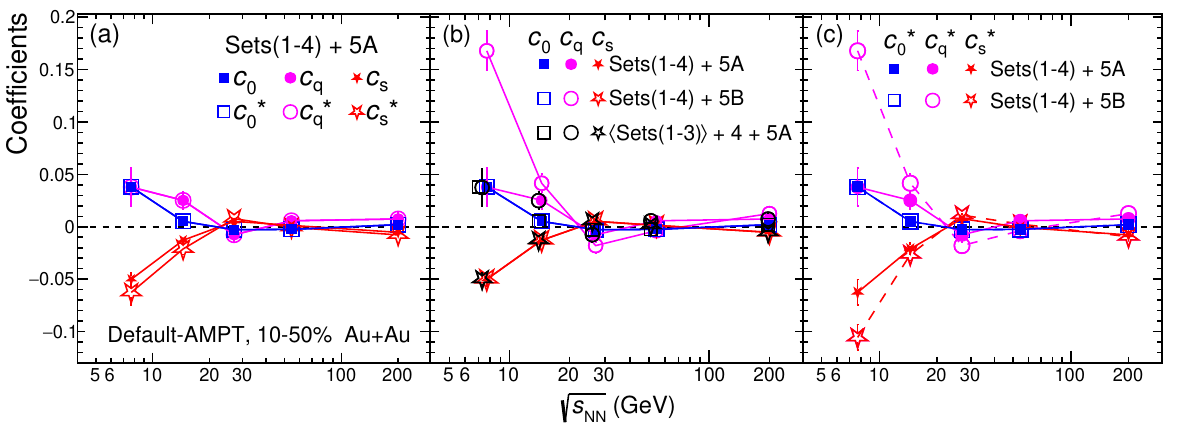} 
\caption{Comparisons of the coefficients extracted from AMPT results
  with the 5-set method as functions of colliding energy: (a) 
  $c_0,c_q,c_S$ compared with $c_0^\star,c_q^\star,c_S^\star$ from
  sets 1 to 4 and 5A, (b) $c_0,c_q,c_S$ values, and 
(c) $c_0^\star,c_q^\star,c_S^\star$ values from sets 1 to 4 and 5A
compared with those from sets 1 to 4 and 5B. Panel (b) also shows the
coefficients from the 3-set method (black symbols).}
\label{fig4}
\end{figure*} 

In Fig.~\ref{fig4}, we compare the coefficients 
extracted from the AMPT model results for semi-central
Au+Au collisions versus the colliding energy. 
Figure~\ref{fig4}(a) compares $c_0,c_q,c_S$ in Eq.\eqref{fitqud}
(filled symbols) with $c_0^\star,c_q^\star,c_S^\star$ in
Eq.\eqref{fitq} (open symbols)  extracted with the 5-set method using 
sets 1 to 4 and 5A. 
We confirm the relations of Eqs.\eqref{cqud3set}-\eqref{cq3set} 
in that fitting the data versus $\Delta q_{ud}$ or $\Delta q$ does not 
affect the $c_0$ and $c_q$ coefficients but gives different $c_S$ values. 
We also see that the coefficients here exhibit a clear energy
dependence, especially at low energies. In particular, at 7.7GeV the
nonzero intercept $c_0$ indicates the breaking of the coalescence
sum rule; as a result, one cannot trust Eq.\eqref{eqcoeff} and
interpret the $c_q$ and $c_S$ coefficients as quark-level $\vvp$
differences there. 

Although there are only five independent hadron sets for this study,  
they can be written in different
combinations~\cite{referAshik,STAR:2023wjl}. For example, one can 
choose them as sets 1 to 4 and set 5B (instead of 5A).  
The corresponding coefficients are shown in Fig.~\ref{fig4}(b) 
for $c_0,c_q,c_S$ and in Fig.~\ref{fig4}(c) for
$c_0^\star,c_q^\star,c_S^\star$, in comparison with those extracted
from sets 1 to 4 and set 5A. We see in Fig.~\ref{fig4}(b)  that the
$c_q$ value depends on the choice of the five sets, while $c_0$ and
$c_S$ values do not. This is expected from Eq.\eqref{cqud3set}, which
shows that set 5 only affects the $c_q$ value.  We also show in
Fig.~\ref{fig4}(b) the coefficients  extracted with the 3-set method
of Eq.\eqref{cqud3set} for hadron sets 1 to 4 and set 5A; they are
essentially the same as those extracted with the 5-set method. Note
that, since hadron set 5B in Table~\ref{table1} is a combination of
set 1 and set 5A, the difference in the $c_q$ value 
from using set 5B and that from using set 5A is given by (three times)
the $\Delta \vvp$ value of set 1, which is shown  in Fig.~\ref{fig2}
to be nonzero at low energies. In Fig.~\ref{fig4}(c), we see that both 
the $c_q^\star$ and $c_S^\star$ values depend on the choice of using
set 5A or 5B.  This is consistent with the expectations of
Eq.\eqref{cq3set}, and the nonzero differences are again due to the
nonzero  $\Delta \vvp$ of set 1 (which would be zero if the
coalescence sum rule were exact). Therefore, getting different
coefficient values from different choices of five independent hadron
sets, like a nonzero $c_0$ value, indicates the breaking of the
coalescence sum rule. 

\section{Conclusions}
\label{conclusions}

In this study, we start from the coalescence sum rule and derive the
relations between the rapidity-odd directed flows ($\vv$) of different
hadron sets. Following earlier studies, we consider seven species of
produced hadrons (those without $u$ or $d$ constituent quarks):
$K^-,\phi,\bar p, \bar \Lambda, {\bar \Xi}^+, \Omega^-$, and ${\bar
  \Omega}^+$,  where the two sides of each hadron set have the same
total number of $\bar u$ and $\bar d$ quarks and the same total number
of $s$ and $\bar s$ quarks after including the weighting factors.  
Earlier studies have proposed that a nonzero directed flow difference 
($\Delta \vv$) between the two sides of the hadron sets, especially a
dependence on the electric charge difference $\Delta q$, means the
breaking of the  coalescence sum rule and would indicate the effect of
the electromagnetic fields. Here we show that the coalescence sum rule
only leads to zero $\Delta \vv$ for a hadron set if its two sides
have identical constituent quark content (or equivalently if $\Delta
q=\Delta S=0$). In general, $\Delta \vv$ depends linearly on both
$\Delta S$ and $\Delta q$, or on both $\Delta S$ and $\Delta q_{ud}$
(the electric charge difference in $\bar u$ and $\bar d$ constituent
quarks). The same is true for $\Delta \vvp$, the difference of the
$\vv$ slopes around mid-rapidity ($\vvp$).   
For $\Delta \vvp$, the coefficient $c_q$ for its $\Delta q_{ud}$
dependence reflects the $\bar d$ and $\bar u$ quark $\vvp$ difference,
while the coefficient $c_S$ for its $\Delta S$ dependence reflects
half the $\bar s$ and $s$ quark $\vvp$ difference. 

Since there are only five independent such hadron sets, there will be
five independent $\Delta \vvp$ data points from the measurement of a
given collision system. We propose to fit the data points with
a two-dimensional plane in the functional form of $c_0+c_q \Delta
q_{ud} + c_S \Delta S$ to extract the three coefficients, where a
nonzero intercept $c_0$ indicates the breaking of the coalescence sum
rule. In the 5-set method, one simply fits the five data points with
this function.  In the more elegant 3-set method, we combine the data
points from the three sets at $\Delta  q_{ud}=\Delta S=0$ into one and
then obtain the coefficients analytically.  
We have also used results from the default version of the AMPT model
for mid-central Au+Au collisions at various energies to
demonstrate the extraction methods. 
The 5-set method and the 3-set method are shown to extract essentially
the same coefficients. 
In addition, we show that the extracted coefficients may depend on the
choice of the five independent hadron sets, and getting different
coefficients from different choices indicates the breaking of the 
coalescence sum rule. This work provides the baseline relations for the $\vv$ 
difference of various hadron sets from the coalescence sum
rule. Further work is needed to consider the possible effect of the
electromagnetic fields. 

\section*{Acknowledgments}
K.N. thanks Prof. Bedangadas Mohanty for providing computational
facilities and hospitality to stay in NISER during the working of this
paper. S.S. is supported in part by the National Key Research and
Development Program of China under Grant No. 2022YFA1604900, 
2020YFE0202002, and the National Natural Science Foundation of China 
under Grant No. 12175084, 11890710 (11890711). 
Z.-W.L is supported by the National Science Foundation under Grant
No. 2012947 and 2310021.

\bibliography{dv1}

\end{document}